\documentclass[aps,prb,twocolumn,superscriptaddress,longbibliography]{revtex4-1}  
\usepackage{graphicx}  
\usepackage{hyperref}
\usepackage{amsmath}
\usepackage{color}

\begin{document}

\title{Antichiral states in twisted graphene multilayers}

\author{M. Michael Denner}
\affiliation{Institute for Theoretical Physics, ETH Zurich, 8093 Zurich, Switzerland}
\affiliation{Department of Physics, University of Zurich, Winterthurerstrasse 190, 8057 Zurich, Switzerland}
\author{J. L. Lado}
\affiliation{Department of Applied Physics, Aalto University, 00076 Aalto, Espoo, Finland}
\affiliation{Institute for Theoretical Physics, ETH Zurich, 8093 Zurich, Switzerland}
\author{Oded Zilberberg}
\affiliation{Institute for Theoretical Physics, ETH Zurich, 8093 Zurich, Switzerland}

\date{\today}

\begin{abstract}
The advent of topological phases of matter revealed a variety of observed boundary phenomena, such as chiral and helical modes found at the edges of two-dimensional (2D) topological insulators. Antichiral states in 2D semimetals, i.e., copropagating edge modes on opposite edges compensated by a counterpropagating bulk current, are also predicted, but, to date, no realization of such states in a solid-state system has been found. Here, we put forward a procedure to realize antichiral states in twisted van der Waals multilayers, by combining the electronic Dirac-cone spectra of each layer through the combination of the orbital moir\'e superstructure, an in-plane magnetic field, and inter-layer bias voltage. In particular, we demonstrate that a twisted van der Waals heterostructure consisting of graphene/two layers of hexagonal boron nitride [(hBN)$_2$]/graphene will show antichiral states at in-plane magnetic fields of 8 T, for a rotation angle of 0.2$^{\circ}$ between the graphene layers. Our findings engender a controllable procedure to engineer antichiral states in solid-state systems, as well as in quantum engineered metamaterials.
\end{abstract}

\maketitle

\section{Introduction}

Dirac materials have sparked vast interest in recent years, as their unique
electronic properties offer a controllable setting with which to realize new states of matter~\cite{Wehling_2014,RevModPhys.81.109}, as well as engineer topological phenomena~\cite{Hatsugai_2011,RevModPhys.82.3045}. A paradigmatic example of a two-dimensional (2D) Dirac material is
graphene~\cite{Choi_2010, Novoselov_2004}, whose spectrum exhibits Dirac-like 
cones at the corners (valleys) $K$ and $K'$ of its hexagonal Brillouin zone. The
two inequivalent $K$ and $K'$ valleys have opposite chiralities with associated quantized
Berry phase~\cite{RevModPhys.81.109}. Correspondingly, in a finite system of zigzag
termination, these valleys are connected by topological dispersionless edge
states~\cite{Brey_2006, Delplace_2011}. This flat edge band has
been observed experimentally in a variety of systems~\cite{Tao2011,Magda2014,Wang2016,Grning2018}.
Interestingly, such Dirac materials can be considered as 
ideal starting points for realizing other exotic surface modes~\cite{Brey_2006,Haldane_1988,PhysRevLett.95.226801,Lado2015,Colom_2018}, by 
introducing proper perturbations to the Dirac 
cones.

A paradigmatic example of the
versatility of the Dirac system
consists of 
breaking time-reversal symmetry in the
honeycomb lattice and opening up
a valley-dependent mass~\cite{Haldane_1988}. In this
situation, a topologically nontrivial bulk gap opens at the Dirac points,
and the above-mentioned flat edge band develops into
the chiral subgap modes of a Chern insulator, where the latter are dispersive and counterpropagating on opposite edges of the 2D material
~\cite{Thouless_1982,Oka_2009}. Even though
this state has not been observed in  solid-state graphene, 
its proposal led to generalizations in other materials, such as
magnetically doped topological
insulators~\cite{Yu2_2010,Chang_2015} and twisted bilayer
graphene~\cite{Sharpe_2019,Serlin_2019}, as well as in engineered emulation of Chern physics in cold atoms~\cite{Esslinger_2014,Flaschner_2016} and photonics~\cite{Rechtsman_2013,Ozawa_2019}. 

A direct generalization of such Dirac spectrum engineering
 includes a perturbation with a spin- and valley-dependent mass~\cite{PhysRevLett.95.226801}, which gives rise to quantum spin Hall insulators. Here, the former flat edge band states develop into helical subgap modes~\cite{Konig_2008}, i.e., the bulk spectrum is gapped, but at each edge a pair of counterpropagating states with opposite spins appear. In graphene,  spin-orbit coupling
creates a rather small topological gap, making observation of such physics challenging. Analogous physics, however, appear in more complex materials, such as monolayer 
tungsten ditelluride~\cite{Wu_2018}. 

Interestingly, antichiral edge modes appear in systems where the bulk spectrum is not gapped~\cite{Colom_2018}.  Here, on opposite edges of the 2D system, the modes propagate in the same direction, compensated by oppositely dispersive semi-bulk-modes. Such antichiral states are generated by introducing a valley-dependent energy
shift to the Dirac cones, which can be mathematically engineered
using tailored long-range hopping
amplitudes~\cite{Colom_2018,PhysRevB.99.161404,PhysRevB.101.195133}. 
Such hopping is  
not present in real graphene, and even though antichiral states
are technically viable~\cite{Chong_2020}, solid-state realizations of them have not been
found so far.

In this paper, we put forward a mechanism to create antichiral states based on 
graphene multilayers and applied electromagnetic fields. 
The system's additional layer degrees of freedom bring forth antichiral states in a realistic solid-state platform through a combination of an applied 
in-plane magnetic field and interlayer bias voltage. Thus Dirac cones of different 
layers are selectively modified, so that copropagating edge
modes in opposite edges manifest.
We first illustrate this idea in aligned bilayer sheets,
providing a mechanism that can be readily applied to 
cold atom setups~\cite{Goldman_2015,Ozawa_2016,Tarruell2012} and artificial
Dirac systems~\cite{Wang2017,Gomes2012}.
We then extend our proposal to 
a graphene/two layers of hexagonal boron nitride [(hBN)$_2$]/graphene twisted multilayer,
demonstrating that
the increased effective lattice constant 
allows for the creation of antichiral states 
at a magnetic field of 8 T for a 0.2$^{\circ}$ twist rotation.
Our proposal puts forward a viable scheme by 
which to realize antichiral states in twisted 
van der Waals systems, opening up future experimental studies 
of antichiral systems.

The paper is organized as follows: In Sec. \ref{sec:2}, we start by
illustrating the fundamental mechanism on $AB$-stacked bilayer graphene, exploring 
the impact of in-plane magnetic fields and interlayer bias voltage 
on both the bulk and edge modes of the system. Specifically, we show that
the in-plane magnetic field shifts the Brillouin zones of the two layers relative to 
one another, such that Dirac cones of different layers can couple in  reciprocal space. 
In this setting, the added interlayer bias voltage tilts the relative energies of the 
two cones, revealing antichiral edge states in a finite system. In Sec.~\ref{sec:3}, 
we demonstrate that a similar phenomenology happens in a twisted bilayer system, where the
emergence of the moir\'e length dramatically reduces the required magnetic field.
Specifically, we show that a graphene/(hBN)$_2$/graphene twisted multilayer
provides a feasible van der Waals system for the creation of antichiral states.
Finally, in Sec. \ref{sec:4}, we summarize our results and provide an outlook to our findings.

\section{Antichiral states in $AB$-stacked graphene bilayers}
\label{sec:2}

To illustrate how antichiral states can be engineered in a stacked $AB$ graphene
bilayer, let us first briefly summarize the electronic properties of graphene
monolayer and aligned bilayers.
Single-layer graphene is a hexagonal two-dimensional material with two atoms 
(denoted $A$ and $B$) per unit cell. As the atoms are identical (carbon), its 
energy spectrum exhibits massless Dirac-cone band touchings at the six corners 
of its hexagonal Brillouin zone. The cones appear in pairs with opposite chiralities 
at the so-called $K$ and $K'$ valleys~\cite{RevModPhys.81.109}. At a proper termination of 
the graphene system, e.g., at a zigzag termination, flat edge bands spectrally connect 
the two valleys. Using a coupled-wire description of the 2D system, these edge bands 
can be understood to be made of uncoupled 0D bound modes of 1D topological
insulators~\cite{Delplace_2011}. 
Including a sublattice-dependent second-neighbor
hopping~\cite{Colom_2018},
the Dirac cones can be shifted in energy, turning
the topological flat bands into antichiral states. However, real graphene
monolayers do not host such long-range hopping, and the Dirac cones are located at the
same energy. As a result, such a valley-dependent energy shift must
be artificially engineered.

\begin{figure}
\centering
\includegraphics[width=.5\textwidth]{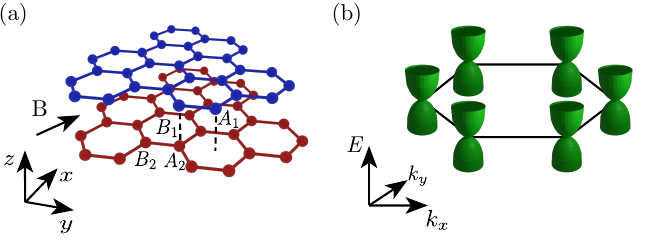}
\caption{$AB$-stacked bilayer graphene. (a) Real-space lattice of $AB$-stacked bilayer graphene, consisting of two graphene sheets that are shifted with respect to each other, while being aligned only at two points of the unit cell. (b) Corresponding schematic low-energy spectrum in the hexagonal Brillouin zone. The combination of the two graphene sheets results in a quadratic low-energy band structure, appearing at the six corners of the first Brillouin zone.}
\label{fig:1}
\end{figure}

Graphene bilayers can provide a possible platform to engineer valley-dependent energy shifts.
They consist of two coupled graphene monolayers that can be arranged in
three main configurations: $AA$, $AB$, and twisted stacking. In the following, we start by considering
$AB$-stacked graphene, in which an atom of the $B_1$ sublattice is directly situated above
an atom of the $A_2$ sublattice, where 1 and 2 denote top and bottom layers, respectively.
The remaining $A_1$ and $B_2$ lie at the center of the honeycombs formed by the
complementary layer; see Fig. \ref{fig:1}(a). The unit cell of the $AB$-stacked bilayer
therefore contains four atoms. Similarly to single-layer graphene, the corresponding
low-energy spectrum exhibits six band touching points divided into two inequivalent
valleys. Contrary to the graphene monolayer, the spectrum in these valleys is not 
linear, but displays a quadratic dispersion; see Fig.~\ref{fig:1}(b)~\cite{McCann_2013,ROZHKOV_2016}. We can intuitively understand this difference: In the absence of interlayer tunnel coupling, the system has
four Dirac cones (two per layer and two per valley). The interlayer coupling hybridizes
the $B_1$ and $A_2$ orbitals and repels states from half filling, leading to high-energy
bands. The states localized at the uncoupled $A_1$ and $B_2$ sites remain uncoupled and
generate the low-energy quadratic touching points. In this bilayer case, Dirac cones
between different layers can be shifted in energy simply by applying
a bias between the layers, yet the already existing hybridization between
cones gives rise to a gap opening~\cite{PhysRevLett.99.216802} and the emergence of a quantum
valley Hall state~\cite{PhysRevLett.107.256801}.
In the following, we will show how this gapping out
of all the Dirac cones is avoided by creating momentum shifts
with applied in-plane magnetic fields.

\begin{figure}
\centering
\includegraphics[width=.5\textwidth]{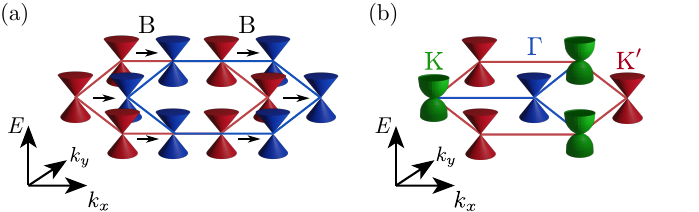}
\caption{$AB$-stacked graphene subject to an in-plane magnetic field. Schematic depiction of the low-energy band structure in the first Brillouin zone for (a) a small in-plane magnetic field $B$ and (b) a large $B$ that hybridizes the cones once more. The Lorentz force acting on electrons tunneling between the two graphene layers causes a momentum separation between the lower (red) and upper (blue) layer spectra [arrows in (a)], thus effectively decoupling the layer Dirac-cone spectra from one another. A strong in-plane magnetic field is able to merge Dirac cones by re-combining different valleys, thus selectively producing quadratic spectra at the merged points, while separating the band structures of the layers in other parts of the Brillouin zone.}
\label{fig:2}
\end{figure}

Applying an in-plane magnetic field between the two graphene layers modifies the
quasimomentum of each graphene layer and is described by the minimal coupling $\mathbf{p} \rightarrow \mathbf{p} -  \mathbf{A}$. The vector potential $\mathbf{A}$
incorporates the magnetic field, which can, for a general in-plane field with orientation
$\phi$, be written as $\mathbf{B} = B \left(\sin\left(\phi\right), \cos\left(\phi\right),  0 \right)^{\text{T}}$ with $\mathbf{A} = B \left(\cos\left(\phi\right)z,
-\sin\left(\phi\right)z, 0\right)^{\text{T}}$. Choosing the coordinate origin to coincide
with a lattice site of the lower layer, we obtain that only the quasimomentum in the upper
layer is affected by the in-plane magnetic field. This implies that the in-plane electron
momentum in the upper layer is changed by the external field according to $\mathbf{p}
\rightarrow \mathbf{p} + \boldsymbol{\Delta p}$ with $\boldsymbol{\Delta p} = B d
\left(-\cos\left(\phi\right), \sin\left(\phi\right), 0 \right)^{\text{T}}$, where $d$ is the
distance between the two layers. Equivalently, this can be understood as a momentum kick 
by a Lorentz force acting on electrons tunneling between the two
layers~\cite{Pershoguba_2010}. This causes the quadratic spectrum to separate back into
Dirac cones of the upper and lower layer~\cite{Pershoguba_2010,Donck_2016}, which are now
shifted in momentum with respect to each other; see Fig.~\ref{fig:2}(a). The separation of
the Brillouin zones of the two layers is proportional to the strength $B$ and depends 
on the orientation $\phi$ of the applied magnetic field.

As the Dirac cones are shifted with respect to each other, the possibility arises to 
merge them once more at specific angles and field strengths. An interesting situation occurs
for a field strength of $B = \frac{1}{a d} \frac{4 \pi}{3 \sqrt{3}}$, where $a$ is the
lattice constant and $d$ is the layer spacing. Here, the field strength corresponds to the
distance between opposite valleys in the hexagonal Brillouin zone. Thus only one pair of
valleys from the two layers is merged, whereas the other valley pair remains separate; see
Fig.~\ref{fig:2}(b). 
Correspondingly, the merged valleys exhibit quadratic dispersion band touchings as in 
the $AB$-stacked case, while the valleys that are not merged exhibit linear Dirac cones 
as in a monolayer graphene. At the corners of the hexagonal Brillouin zone, the latter 
are formed by the lower layer that is unaffected by the magnetic field. At the
$\Gamma$ point, the shifted upper layer's Dirac cone appears.

\begin{figure}
\centering
\includegraphics[width=.47\textwidth]{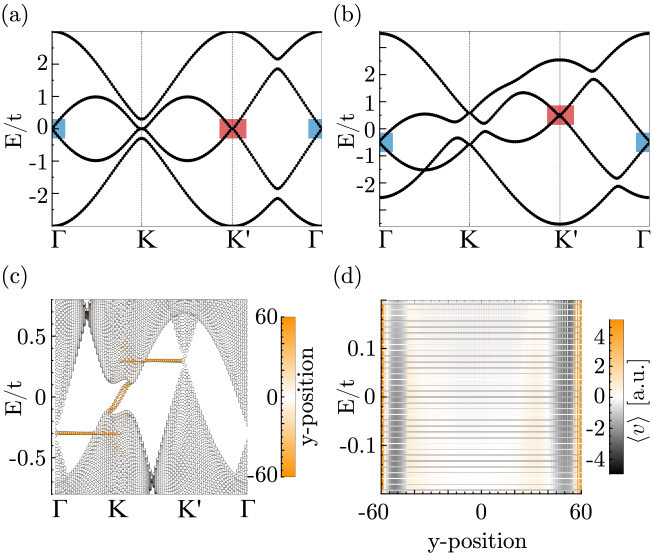}
\caption{Band structure of $AB$-stacked bilayer graphene. (a) The spectrum with an in-plane magnetic field that merges Dirac cones from different layers at the $K$ points (quadratic touching), and leaves linear uncoupled Dirac cones at the $\Gamma$ and $K'$ points (marked by blue and red squares); cf.~Fig.~\ref{fig:2}. (b) Since the quadratic touching points are formed by states originating from different layers, an interlayer voltage bias gaps them. Moreover, the bias voltage shifts in energy the Dirac cones of the upper and lower layers relative to one another. (c) This procedure reveals antichiral edge states when considering a finite system of 20 unit cells with zigzag termination, with the colormap indicating the position of eigenstates along the finite dimension. (d) The antichiral states propagate in the same direction along the boundary (orange) compensated by a bulk current (black) of opposite direction close to the edge, as can be seen by the spatially resolved group velocity.} 
\label{fig:3}
\end{figure}

The effect of the magnetic field can be incorporated into a tight-binding description of the system using Peierls's substitution, i.e., by modifying the hopping amplitudes as
\begin{equation}
    t_{ij} \rightarrow t_{ij}\ e^{i \Phi_{ij}} = t_{ij}\ e^{i \int_{\mathbf{r}_i}^{\mathbf{r}_j} \mathbf{A} \cdot d\mathbf{r}}\,,
\label{eq:hop}
\end{equation}
where $\mathbf{A}$ is the vector potential, $\mathbf{r}_{i,j}$ are the positions of the atoms in the material, and $t_{ij}$ is the bare hopping amplitude between electron valence orbits on atoms $i$ and $j$. We, thus, write down the tight-binding Hamiltonian for $AB$-stacked graphene~\cite{McCann_2013,ROZHKOV_2016} in the presence of an in-plane magnetic field and obtain the corresponding spectrum; see Fig.~\ref{fig:3}(a). Similar to the effective low-energy description in Fig.~\ref{fig:2}, the in-plane magnetic field  clearly separates the band structures of the two layers in certain areas of reciprocal space. In the nonmerged valley at the $\Gamma$ point, the spectrum originates from isolated monolayers, whereas the merged valley exhibits a quadratic dispersion caused by interlayer coupling.

By applying a bias voltage between the stacked layers, the energies of the states in each layer are shifted with respect to one another (inducing layer polarization), and a gap opens up at the quadratic touching points~\cite{McCann_2006, McCann_20061, Hongki_2007}. As a consequence, in standard $AB$-stacked graphene with $B=0$,  the system is fully gapped, as the states forming the quadratic touching points are split off. Introducing an in-plane magnetic field $B\neq0$, a similar behavior is obtained at the merged valley, where the quadratic dispersion is replaced by a band gap, which is tunable by the strength of the induced layer polarization; see Fig.~\ref{fig:3}(b). At the same time, since the in-plane field decoupled the spectra of the two layers at the $\Gamma$ and $K'$ points, the full system is not gapped, but instead, the Dirac cone originating from the upper layer is shifted opposite in energy to the Dirac cone of the lower layer. In turn, this means that the system is no longer a semimetal, but it becomes a conductor, as there is a finite density of states crossing at any filling of the material. 

Interestingly, at half filling, the bulk becomes conducting with a bulk current that is
compensated by copropagating edge modes. The latter antichiral states are revealed in a
finite system; see Fig.~\ref{fig:3}(c). A special feature of graphene is that these edge
modes emerge only in certain geometries, for example, in bilayer graphene nanoribbons with
zigzag edges~\cite{Nakada_1996,Fujita_1996,Yao_2009,Delplace_2011}. Specifically, in our
case, a finite bilayer system with zigzag termination [Fig.~\ref{fig:3}(c)] shows flat edge
bands connecting the Dirac cones of each layer that are then forced to disperse due to the shifted
energy between the layers. To ensure charge neutrality, this flow of charge carriers at the
boundary is compensated by a bulk current flowing in the opposite direction; see
Fig.~\ref{fig:3}(d). This is reflected in the bulk bands crossing half filling at the
not-merged valleys at the $\Gamma$ and $K'$ points. Note that in a perfect armchair geometry, antichiral edge states are hidden by the projection of both graphene valleys onto the same momentum.

We thus obtain a scenario that generates antichiral states in a realistic system, using the
tunability offered by the application of magnetic and electric fields on bilayer systems. Note
that the key ingredient of shifted Dirac cones that drag in energy the topological flat edge bands
manifests here in similitude to the case of the modified Haldane model~\cite{Colom_2018}.
Furthermore, we can reverse the propagation direction of these antichiral states either by
changing the sign of the layer polarization or by using an opposite field to merge the Dirac
cones at the $K'$ valley. 
However, the in-plane magnetic field required to shift the stacked graphene Brillouin zones with respect to each other for a real graphene bilayer corresponds to an unrealistically large value,
$
   B = \frac{1}{a d} \frac{4 \pi}{3 \sqrt{3}} = 19.6\ \mathrm{kT},
$ where we converted to real units by taking $a = 2.46\
\mathring{A}$~\cite{McCann_2013} and $d = 3.3\ \mathring{A}$~\cite{Pershoguba_2010}. 
Nevertheless, it is important
to emphasize that such large effective fields can be achieved
in quantum engineered systems, where synthetic
gauge fields can be induced~\cite{Bloch_2008,Ozawa_2019}. 
As in this paper we are ultimately interested in showing how
to create antichiral states in  solid-state materials, in the following, we
show how an analogous mechanism can be realized in twisted bilayer graphene, dramatically
lowering the required magnetic fields.

\section{Antichiral states in twisted bilayer graphene}
\label{sec:3}
\begin{figure}
\centering
\includegraphics[width=.45\textwidth]{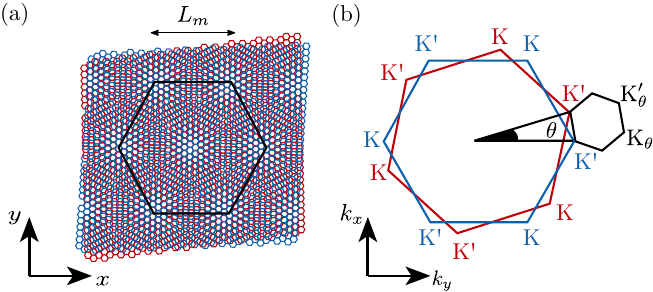}
\caption{Twisted bilayer graphene. (a) Real-space moir\'e pattern formed by the relative rotation of two graphene layers (red and blue mark atoms in the lower and upper layers, respectively). (b) Corresponding Brillouin zones of the lower (red) and upper (blue) layer rotated by angle $\theta$. The superstructure in real space of length $L_m$ induces a much smaller Brillouin zone in reciprocal space.} 
\label{fig:4}
\end{figure}

In the previous section, we saw that the $AB$-stacked bilayer graphene exhibits anti-chiral
states once two coupled Dirac cones are shifted in energy relative to one another. Yet, the
in-plane magnetic field required for selective coupling of these Dirac cones was
too large, a feature simply related to the lattice
constant and layer spacing associated with such a momentum space translation. 
Hence, in this section, we move to investigate a similar scenario in twisted bilayer
graphene, in which the emergence of a new moir\'e length will dramatically
lower the magnetic field required. 
In such a system, the two graphene layers are not perfectly aligned, but form a 
relative angle, creating a so-called moir\'e superstructure~\cite{Rozhov_2017}. 
This causes an alternating pattern of $AB$, $AA$, and $BA$ stackings, resulting in a
supercell up to 1000 times larger than in a graphene monolayer; 
see Fig.~\ref{fig:4}(a). Such moir\'e patterns are observed with scanning 
tunneling spectroscopy~\cite{Zhao_2011}, and the period is described 
as~\cite{Rozhov_2017}
\begin{equation}
L_M = \frac{a}{2 \sin \frac{\theta}{2}} \approx \frac{1}{\theta}\,,
\end{equation}
such that the size of the unit cell scales roughly as $\frac{1}{\theta}$ for 
small twisting angles $\theta$. The corresponding first Brillouin zone is 
again hexagonal but considerably smaller than that of a normal graphene layer; 
see Fig.~\ref{fig:4}(b)~\cite{Rozhov_2017}. Correspondingly, this yields a reduced inter-Dirac-cone reciprocal distance
\begin{equation}
\Delta K = \vert \mathbf{K}_{\theta}-\mathbf{K'}_{\theta} \vert = \frac{4 \pi}{3 a} \sqrt{2} \sqrt{1- \cos \theta}\,.
\end{equation}
Hence, for small twisting angles, the twist dramatically reduces the distance between the Dirac valleys, i.e., this allows for much smaller in-plane magnetic fields necessary for the  generation of selective Dirac-cone coupling and the appearance of antichiral states; cf.~section above.

\begin{figure*}
\centering
\includegraphics[width=1.\textwidth]{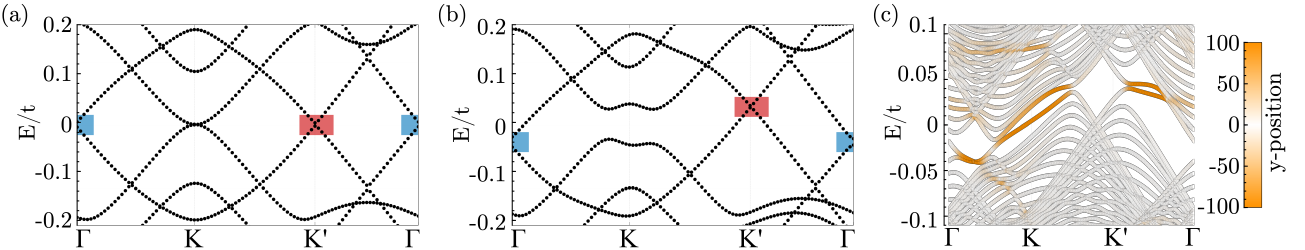}
\caption{Band structure of twisted bilayer graphene in the antichiral regime. 
(a) In the presence of an in-plane magnetic field that selectively merges Dirac cones 
from different layers (cf.~Figs.~\ref{fig:2} and~\ref{fig:3}), the in-plane magnetic 
field reduces the number of Dirac cones by creating quadratic touching points in the 
merged valley. (b) An applied interlayer bias voltage opens a gap at the quadratic 
touching points and shifts the unpaired Dirac cones from the different layers relative 
to one another (red and blue squares). (c) Antichiral edge states appear when 
considering a finite system, with the colormap indicating the position of the 
eigenstates along the finite dimension.
It is observed that the states localized on opposite edges propagate in the
same direction, as expected from antichiral modes.
This regime is achieved in the twisted graphene/(hBN)$_2$/graphene
heterostructure for a rotation angle of $0.2^{\circ}$ and an in-plane magnetic field
of 8 T.
} 
\label{fig:5}
\end{figure*}

The tight-binding Hamiltonian describing the twisted bilayer graphene in the presence of in-plane magnetic fields can be written as~\cite{PhysRevLett.99.256802,PhysRevB.92.075402,PhysRevB.82.121407,PhysRevLett.123.096802}
\begin{align}
H =& \sum_{\langle i,j \rangle} t\left(\Phi, \mathbf{r}_i, \mathbf{r}_j \right) c_i^{\dagger} c_j^{} + \sum_{i,j} \hat{t}_{\perp} \left(\Phi, \mathbf{r}_i, \mathbf{r}_j \right)c_i^{\dagger} c_j^{} \notag \\ 
&+ \Delta \sum_i \tau_z^{ii} c_i^{\dagger} c_i^{}\,.
\end{align}
The first term describes  nearest-neighbor $\langle i,j \rangle$ hopping within the layers,
with the hopping amplitude $t$ modified by Peierls's substitution (\ref{eq:hop}). The hopping
between the twisted layers $\hat{t}_{\perp} \left(\Phi, \mathbf{r}_i, \mathbf{r}_j \right)$
depends on the relative distance between the atoms on different layers, and has its maximum
value $t_{\perp}$ for perfect stacking. The last term describes the bias-voltage induced
layer polarization. We denote  $\Phi = B a d$, to be the flux piercing the interlayer
plaquette. As a reference, in graphene the parameters are~\cite{McCann_2013} $t = 3~\text{eV}$ and $t_{\perp} = 300~\text{meV}$.
Similar to the aligned $AB$-stacked case in the previous section, the in-plane magnetic 
field causes the energy spectrum to separate into two cones per valley, which are shifted 
by the applied field~\cite{Kwan_2019}. This means that their separation can be controlled 
by the angle $\phi$ and strength $B$ of the field. As a reference, the 
distance between the cones can be bridged by a field 
of $\Phi \approx 0.02$ in natural units, corresponding to a magnetic 
field of $B = 162.2~\text{T}$, which is beyond experimentally feasible values.

On the bright side, by exploiting twist engineering, we have dramatically reduced the
magnetic field required to merge the Dirac points in twisted graphene
bilayers. Nevertheless, the field strength required is still too large for
a feasible realization of antichiral states
in twisted bilayer graphene at angles above 1$^{\circ}$.
A simple recipe to bring down the required fields is to increase
the flux associated to a certain rotation angle or to decrease the interlayer hopping
to conserve the Dirac cones at smaller rotation angles. Luckily, both schemes can
be achieved by including a thin hBN insulating layer between the twisted
graphene layers, where the effective hopping between the layers in the resulting graphene/hBN/graphene heterostructures can be tuned
between 2 to 70 meV when inserting three to one hBN layers.\cite{Valsaraj2016} In particular,
taking two layers of hBN as spacers brings the effective interlayer hopping $t_\perp$
to 20 meV,\cite{Valsaraj2016} while preserving the original Dirac-cone spectra up to rotation angles
of 0.1$^{\circ}$~\cite{PhysRevLett.99.256802,PhysRevB.82.121407,Bistritzer2011} (see Appendix~\ref{ap:1}).
Furthermore, the inclusion of two hBN layers increases the
interlayer distance by approximately a factor of 3~\cite{Valsaraj2016}.
Taking this into account, we obtain that a twisted heterostructure
graphene/(hBN)$_2$/graphene would show antichiral states for a rotation angle
of 0.2$^{\circ}$ and a magnetic field of 8 T, which is an experimentally
achievable regime. This demonstrates that antichiral states can be realistically
engineered in twisted graphene-hBN superlattices, opening up a feasible
solid-state platform for antichiral physics.

We now explicitly show that antichiral states indeed emerge in the effective
model for twisted bilayer graphene. We directly focus on the regime in which
the associated magnetic flux is comparable to the lattice constant, where for computational convenience we use a rescaling trick~\cite{PhysRevLett.119.107201}.
In this situation, the number of Dirac cones 
can be reduced by merging opposite graphene valleys by applying
the required in-plane magnetic flux [Fig. \ref{fig:5}(a)], which
will be 8 T for the graphene/(hBN)$_2$/graphene
heterostructure at a rotation angle
of 0.2$^{\circ}$. In order to realize antichiral
states, the two remaining Dirac cones are shifted in energy relative to one another. This is
again achieved by adding an interlayer bias voltage [Fig. \ref{fig:5}(b)], 
i.e., on-site energies $\Delta$ that 
have a different sign $\tau_z^{ii} = \pm 1$ depending on the corresponding layer. This 
causes the separated Dirac cones of the upper and lower layer to be shifted in energy, 
whereas the quadratic merging point is gapped out. At the gapped merging point, 
antichiral edge states appear. These states are expected to always appear, due to the fact that for a twisted graphene geometry it is impossible to have purely armchair edges.
Since the remaining cones are shifted up and down in energy, bulk states are present at 
the same chemical potential that compensate 
the copropagating edge channels [Fig. \ref{fig:5}(b)]. The system
is no longer insulating, but metallic, with currents along the edges propagating opposite
to the bulk flow [Fig. \ref{fig:5}(c)]. 
To summarize,  antichiral states can  be obtained in a twisted
bilayer graphene system using the same recipe of applying in-plane magnetic and interlayer 
electric fields.

We recall that 
in the $AB$-stacked model in Sec.~\ref{sec:2}, the direction of the antichiral 
states could be reversed either by changing the sign of the gap-opening bias or 
by reversing the field direction. Interestingly, applying such a modification 
in the twisted case shows no change in the propagation direction. 
This is because of the lack of rotational symmetry of the twisted layers. 
Changing the signs of the on-site potentials effectively corresponds to rotating the
graphene sheets around an in-plane axis, thus also changing the direction of the 
currents along the edges. In a normal $AA$ or $AB$ stacking, this produces the same
crystal
structure, with reversed layer polarization. In contrast, in twisted bilayer graphene, 
the angle between the two layers destroys this symmetry, such that a rotation creates 
a new structure, possessing identical current directions. The direction of the currents 
can, however, still be reversed by the time-reversal breaking direction, which corresponds 
to an inversion of the magnetic field. This means merging Dirac cones in a different
 valley. Since these valleys are related by time-reversal symmetry, the 
propagation direction is reversed.

\section{Conclusions}
\label{sec:4}
To summarize, we put forward a procedure to create
antichiral states in graphene multilayer systems, by combining
magnetic fields and interlayer bias. In particular, we have demonstrated
that a twisted graphene/(hBN)$_2$/graphene heterostructure
at 0.2$^{\circ}$ rotation will show antichiral states for in-plane magnetic fields
of 8 T. This fundamental idea consists of engineering
a system hosting two Dirac points that can be shifted in energy
by means of an interlayer bias.
This is achieved by shifting 
the Dirac cones 
by an in-plane magnetic field, 
with the regime being reached when the in-plane magnetic flux
is comparable to the moir\'e length.
Besides a realistic van der Waals solid-state realization, 
we have also proposed a minimal
system consisting of aligned graphene honeycomb lattices
in which antichiral states can be created. Such a minimal scheme can be exploited in cold
atom gases and engineered quantum systems.
Our work therefore marks a promising
step towards the realization and engineering of these special antichiral
states, providing a stepping stone towards further studies in antichiral metals.

\section{Acknowledgments}
We acknowledge financial support from the Swiss National Science Foundation through Grants No. PP00P2\_1163818 and No. PP00P2\_190078. The authors 
are grateful to F.~Goerg, T.~Wolf, and I.~Petrides for helpful suggestions 
and fruitful discussions. J.L.L. acknowledges the computational resources provided by the Aalto Science-IT project
and financial support from
Academy of Finland Projects No. 331342 and No. 336243.

\appendix
\section{hBN layers as insulating spacers and their impact on the low-energy band structure of the twisted bilayer graphene}
\label{ap:1}

\begin{figure}[t!]
\centering
    \includegraphics[width=\columnwidth]{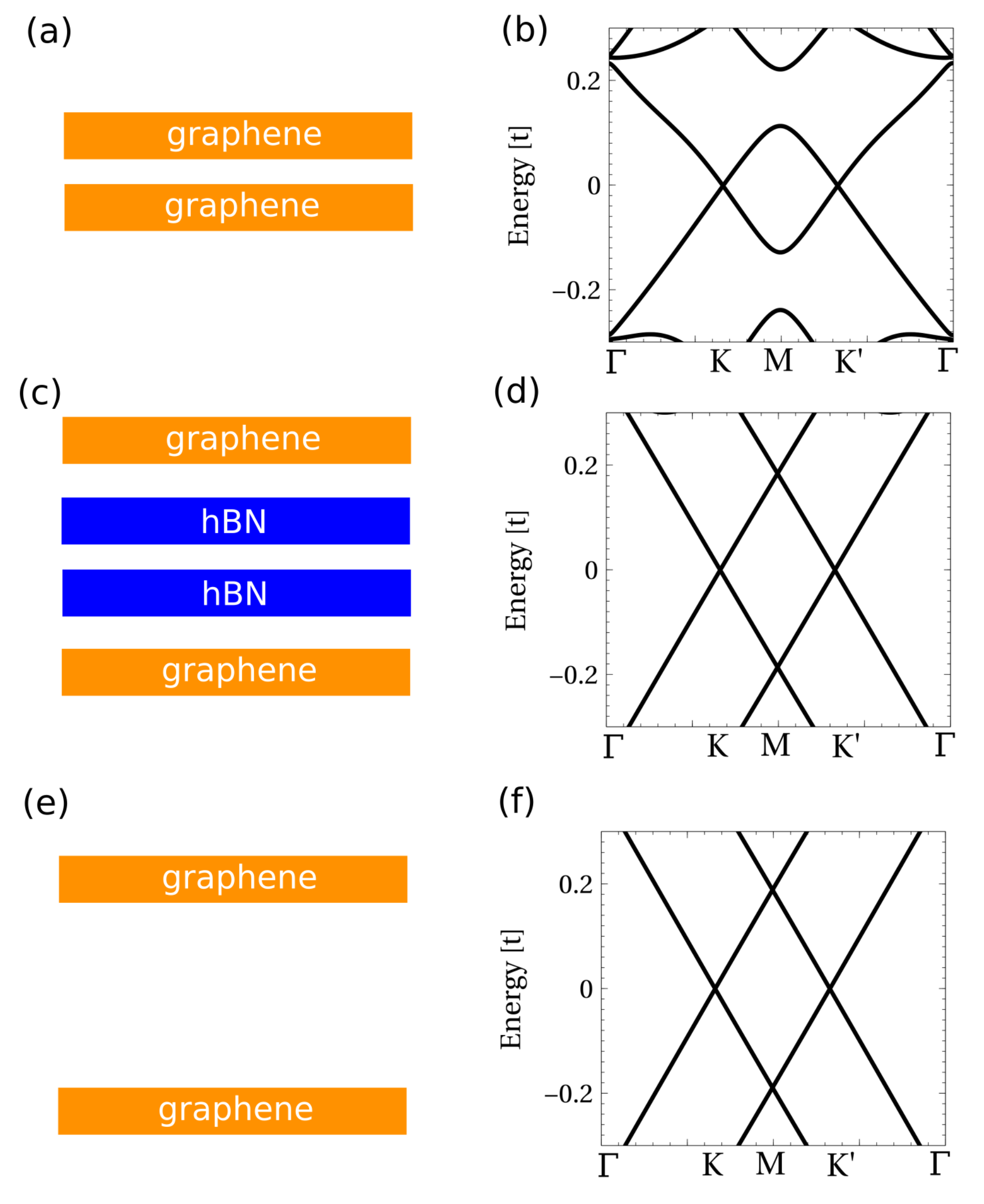}

	\caption{ 
	Sketch (a) and band structure (b)
	of a twisted graphene bilayer with a conventional
	interlayer distance.
	When two twisted layers of hBN are put in between
	the graphene layers (c), the effective interlayer coupling
	becomes weaker (d). By comparing with a structure
	in which the two layers are artificially separated
	yielding a reduced interlayer coupling (e)
	it is observed that the twisted hBN does not create
	additional perturbations (f).
}
\label{fig:SMfig1}
\end{figure}

In this appendix, we illustrate why hBN can be used for encapsulation and for studying graphene devices without strongly impacting the low-energy gapless Dirac structure~\cite{Dean2010,Young2013,Wallbank2015}. In general, adding an hBN monolayer can have an additional impact on the electronic structure of graphene multilayers for specific angles~\cite{Wallbank2015}. For instance, aligning hBN with one of the twisted graphene layers~\cite{Sharpe2019} opens a sizable band gap in the spectrum.
However, the situation is dramatically different when there is a large twist angle between the hBN and twisted graphene bilayers~\cite{Wallbank2015}. In such a situation, the electronic structure remains gapless, due to averaging out of the sublattice imbalance in the moir\'{e} unit cell~\cite{Cao2018,PhysRevX.8.031087}. Correspondingly, a large twist angle between the hBN and the graphene multilayers allows lifting of any nontrivial impact of the hBN on the graphene band structure at low energies and acts as an effective spacer between the graphene layers~\cite{PhysRevB.86.125448}. To demonstrate this, we explicitly compute the electronic structure of a graphene/(hBN)$_2$/graphene twisted heterostructure with a relative angle between the different layers of 6$^{\circ}$, demonstrating that the effect of hBN is to effectively weaken the interlayer coupling yet without impacting the low-energy dispersion. For this, we use an analogous real-space model as in Sec.~\ref{sec:3}, but now taking layers that can have an intrinsic sublattice imbalance $m$
\begin{align}
H =& t \sum_{\langle i,j \rangle}  c_i^{\dagger} c_j^{} + 
	t^{ij}_{\perp}\sum_{i,j}  c_i^{\dagger} c_j^{} \notag \\
	&+ m \sum_{i\in BN} \sigma_z^{ii} c_i^{\dagger} c_i^{}\,,
\end{align}
where the first term describes nearest-neighbor $\langle i,j \rangle$ hopping within the layers, $t_{\perp}$ is the hopping between the layers, and the last term describes the intralayer imbalance between the sublattices modeled with the Pauli $\sigma_z$ matrix. In particular, the hBN layers are modeled by taking $m=1.2t$, which gives rise to a large gap at charge neutrality. We now compute the electronic spectra of a twisted graphene bilayer that has (i) no hBN spacer between layers [Figs.~\ref{fig:SMfig1}(a) and (b)], (ii) two twisted hBN layers as spacers [Figs.~\ref{fig:SMfig1}(c) and (d)], and (iii) a vacuum spacer [Figs.~\ref{fig:SMfig1}(e) and (f)]. It is clearly observed that the two hBN layers as spacers yield a band structure analogous to the one with vacuum spacing, without introducing additional perturbations. This exemplifies that twisted hBN is capable of solely weakening the interlayer coupling in a twisted graphene bilayer, while increasing the interlayer distance. These results were performed with a full atomistic real-space tight-binding model as was Fig.~\ref{fig:5} in Sec.~\ref{sec:3} and thus incorporate all the microscopic details associated with the interatom hopping, without low-energy approximations. 

Interestingly, the same effect created with an hBN spacer could be obtained by a vacuum spacing in suspended graphene devices~\cite{Grushina2013,Nam2017,Nam2018}. These devices would also allow one to create the antichiral states proposed in our work by taking a spacing between twisted layers on the order of 1 nm. Although suspended graphene multilayers have already been fabricated~\cite{Nam2018}, suspended samples are generically more challenging to fabricate than the original graphene/hBN$_2$/graphene devices we propose here.

\bibliography{References}

\end{document}